\documentclass{article}
\usepackage{times,fancyhdr}
\usepackage{cite}
\usepackage{graphicx}

\title{The NLO  ${\cal N} =4$ SUSY BFKL Green function in the adjoint 
representation
} 
\author{G. Chachamis$^1$, A. Sabio Vera$^2$ \\ 
\\
$^1$ Paul Scherrer Institut, CH-5232 Villigen PSI, Switzerland\\
\\
$^2$ Depto. de F{\' 	\i}sica Te{\' o}rica \& Instituto de F{\' \i}sica Te{\' o}rica UAM/CSIC,\\
Universidad Aut{\' o}noma de Madrid,  Cantoblanco E-28049 Madrid, Spain.}

\begin{document} 

\fancyhead[EL,OR]{\thepage}
\renewcommand\headrulewidth{0.5pt}
\addtolength{\headheight}{2pt} 

\maketitle 

We study the solution of the BFKL equation in the adjoint representation for the ${\cal N}=4$ 
SUSY theory at NLO accuracy. Consistency with the large momentum transfer solution obtained by Fadin and Lipatov in~\cite{arXiv:1111.0782} is found. We investigate, for large and small values of the momentum transfer, 
the growth with energy of the Green function, its collinear behaviour and the expansion in azimuthal angle Fourier components.

\section{Introduction}

The Balitsky-Fadin-Kuraev-Lipatov (BFKL) formalism~\cite{BFKL1,BFKL2,BFKL3} has proven to be very 
useful to understand scattering amplitudes in the ${\cal N} = 4$ supersymmetric Yang-Mills theory 
in multi-Regge (MRK) and quasi-multi-Regge (QMRK) kinematics~\cite{Fadin:1998py,Ciafaloni:1998gs}.  In this context, 
corrections to the Bern-Dixon-Smirnov (BDS) iterative ansatz~\cite{Bern:2005iz} were found in this limit for the six-point amplitude at two loops in~\cite{Bartels:2008ce,Bartels:2008sc}. The gluon Green function in the adjoint representation is a key ingredient  in the calculation of these scattering amplitudes and also in the case of an arbitrary number of external legs and loops. We have investigated its all-orders structure at leading order (MRK) in~\cite{Chachamis:2011nz}. In the 
present paper we improve on that study to address the much more complicated case of next-to-leading order (NLO) corrections, solving the corresponding equation presented by Fadin and Lipatov for QMRK in~\cite{arXiv:1111.0782}.  

In Section 2 we provide an iterative representation for the NLO Green function in the adjoint representation 
directly in  rapidity and transverse momentum space. As shown in~\cite{Bartels:2008ce,Bartels:2008sc}, the infrared divergences can be factorized in a simple form, leaving an iterative infrared finite remainder that we numerically investigate, 
with Monte Carlo techniques~\cite{code} in Section 3. We write an iterative representation for the solution to the 
equation for arbitrary momentum transfer $q$. In the limit of large $q$ we obtain agreement with the studies in~\cite{Bartels:2008ce,Bartels:2008sc}. Besides investigating the collinear behaviour of the Green function, we find that the growth with energy of the different Fourier components in the azimuthal angle changes its  structure when going from the large to the small momentum transfer limit.  Finally, we present our Conclusions and scope for future work. 

\section{The adjoint BFKL Green function at NLO in iterative form}

In Refs.~\cite{Fadin:2005zj} the non-forward BFKL equation for QCD 
was presented for a general projection 
of the colour quantum numbers in the $t$-channel, at leading and next-to-leading order. The relevant representation 
for planar ${\cal N} = 4$ SUSY maximally helicity violating amplitudes is the adjoint which, at NLO, was 
calculated in~\cite{arXiv:1111.0782}. The main difference between the octet and the singlet representation is that the latter is infrared finite while the former is not. It is however possible to show that the extra infrared divergencies that appear in the non-singlet representations can be written as a simple overall factor in the gluon Green function~\cite{Bartels:2008ce,Bartels:2008sc,arXiv:1111.0782}. The infrared finite remainder contains infrared divergencies in the real emission and gluon trajectory sectors which have to cancel against each other. Their corresponding divergences can be regularized by a physical cut off in the transverse momentum, which we call $\lambda$. In this way we can then write the ${\cal N}=4$ SUSY BFKL equation in the adjoint representation at NLO in the form
\begin{eqnarray}
\Bigg\{\omega + \frac{\bar \alpha}{2} \left(1-\frac{\zeta_2}{2} {\bar \alpha}\right)
\log{\left( {  {\bf q}_1^2 {\bf q}_1'^2   \over {\bf q}^2 \lambda^2}\right)} 
-\frac{3}{4} \zeta_3 {\bar \alpha}^2
\Bigg\} 
{\cal F}_\omega \left({\bf q}_1,{\bf q}_2;{\bf q}\right) &=&
\delta^{(2)} \left({\bf q}_1-{\bf q}_2\right)\nonumber\\
&&\hspace{-9.5cm}+\int d^2 {\bf k} \Bigg\{
 {{\bar \alpha} \over 4} \left(1-\frac{\zeta_2}{2} {\bar \alpha}\right) 
 {\theta \left({\bf k}^2 - \lambda^2\right) \over \pi {\bf k}^2}
\Bigg(1+{{\bf q}_1'^2 ({\bf q}_1+{\bf k})^2 - {\bf q}^2 {\bf k}^2 \over ({\bf q}_1'+{\bf k})^2 {\bf q}_1^2}\Bigg) + \Phi \left({\bf q}_1,{\bf q}_1 + {\bf k} \right)\Bigg\} \nonumber\\
&&\hspace{-5cm}\times {{\bf q}_1^2 \over \left({\bf q}_1 + {\bf k}\right)^2}
{\cal F}_\omega \left({\bf q}_1+{\bf k},{\bf q}_2;{\bf q}\right). 
\end{eqnarray}
Let us note that we have used the notation ${\bf q}_i' \equiv {\bf q}_i - {\bf q}$, where ${\bf q}$ is the 
momentum transfer and all two--dimensional vectors are represented in bold. We also have 
\begin{eqnarray}
\Phi \left({\bf q}_1,{\bf q}_1 + {\bf k} \right) &=& 	\nonumber\\
&&\hspace{-3cm}{{\bar \alpha}^2 \over 32 \pi} {1 \over {\bf q}_1^2 ({\bf k}+{\bf q}'_1)^2}
\Bigg\{{\bf q}^2 \Bigg[ \ln{\left({{\bf q}_1^2 \over {\bf q}^2} \right)} 
\ln{\left({{{\bf q}_1'}^2 \over {\bf q}^2}\right)} 
+\ln{\left({ \left({\bf q}_1 + {\bf k}\right)^2 \over {\bf q}^2}\right)} 
\ln{\left({\left({\bf q}'_1 + {\bf k}\right)^2 \over {\bf q}^2}\right)} \nonumber\\
&&\hspace{-3cm}+\frac{1}{2} \ln^2{\left({{\bf q}_1^2 \over \left({\bf q}_1 + {\bf k}\right)^2} \right)}
+\frac{1}{2} \ln^2{\left({{\bf q}_1'^2 \over \left({\bf q}_1' + {\bf k}\right)^2} \right)} \Bigg] 
+ \frac{1}{2}  { \left({\bf q}_1^2   \left({\bf q}_1' + {\bf k}\right)^2- {\bf q}_1'^2  \left({\bf q}_1 + {\bf k}\right)^2\right) \over {\bf k}^2} \nonumber\\
&&\hspace{-2.5cm} \times \Bigg[ \ln{\left({{\bf q}_1'^2 \over \left({\bf q}_1' + {\bf k}\right)^2} \right)} 
\ln{\left({{\bf q}_1'^2  \left({\bf q}_1' + {\bf k}\right)^2  \over   {\bf k}^4}\right)}
- \ln{\left({{\bf q}_1^2 \over \left({\bf q}_1 + {\bf k}\right)^2} \right)}
\ln{\left({{\bf q}_1^2  \left({\bf q}_1 + {\bf k}\right)^2  \over   {\bf k}^4}\right)}\Bigg] \nonumber\\
&&\hspace{-3cm}- { \left({\bf q}_1^2   \left({\bf q}_1' + {\bf k}\right)^2+ {\bf q}_1'^2  \left({\bf q}_1 + {\bf k}\right)^2\right) \over {\bf k}^2} \Bigg[\ln^2{\left({{\bf q}_1^2 \over \left({\bf q}_1 + {\bf k}\right)^2} \right)}
+ \ln^2{\left({{\bf q}_1'^2 \over \left({\bf q}_1' + {\bf k}\right)^2} \right)}\Bigg] \nonumber\\
&&\hspace{-3cm} 
+ \Bigg[{\bf q}^2 \left({\bf k}^2 - {\bf q}_1^2 - \left({\bf q}_1 + {\bf k}\right)^2 \right)
+2 {\bf q}_1^2 \left({\bf q}_1 + {\bf k}\right)^2-{\bf q}_1^2 \left({\bf q}_1' + {\bf k}\right)^2
-{\bf q}_1'^2 \left({\bf q}_1 + {\bf k}\right)^2 \nonumber\\
&&\hspace{-2.5cm}
+ { \left({\bf q}_1^2   \left({\bf q}_1' + {\bf k}\right)^2- {\bf q}_1'^2  \left({\bf q}_1 + {\bf k}\right)^2\right) \over {\bf k}^2}  
\left({\bf q}_1^2 - \left({\bf q}_1 + {\bf k}\right)^2\right)\Bigg] 
{\cal I} \left({\bf q}_1^2, \left({\bf q}_1 + {\bf k}\right)^2, {\bf k}^2\right) \nonumber\\
&&\hspace{-3cm} 
+ \Bigg[{\bf q}^2 \left({\bf k}^2 - {\bf q}_1'^2 - \left({\bf q}_1' + {\bf k}\right)^2 \right)
+2 {\bf q}_1'^2 \left({\bf q}_1' + {\bf k}\right)^2-{\bf q}_1'^2 \left({\bf q}_1 + {\bf k}\right)^2
-{\bf q}_1^2 \left({\bf q}_1' + {\bf k}\right)^2 \nonumber\\
&&\hspace{-2.5cm}
+ { \left({\bf q}_1'^2   \left({\bf q}_1 + {\bf k}\right)^2- {\bf q}_1^2  \left({\bf q}_1' + {\bf k}\right)^2\right) \over {\bf k}^2}  
\left({\bf q}_1'^2 - \left({\bf q}_1' + {\bf k}\right)^2\right)\Bigg] 
{\cal I} \left({\bf q}_1'^2, \left({\bf q}_1' + {\bf k}\right)^2, {\bf k}^2\right)\Bigg\}
\end{eqnarray}
with
\begin{eqnarray}
{\cal I} \left({\bf p}^2, {\bf q}^2, {\bf r}^2\right) &=& \int_0^1 \frac{d x}{{\bf p}^2 (1-x)+{\bf q}^2 x- {\bf r}^2 x (1-x)} 
\ln{\left(\frac{{\bf p}^2 (1-x)+{\bf q}^2 x}{{\bf r}^2 x (1-x)}\right)}.
\end{eqnarray}
The function $\Phi$ was calculated by Fadin and Lipatov in~\cite{arXiv:1111.0782}.  

After this regularization 
it is possible to iterate the equation and perform a Mellin transform back into energy (or rapidity) space. The 
final, iterative, representation in transverse momentum and rapidity space for the gluon Green function is 
\begin{eqnarray}
{\cal F} \left({\bf q}_1,{\bf q}_2;{\bf q};{\rm Y}\right) &=& 
\left({{\bf q}^2 \lambda^{2} \over {\bf q}_1^2 {\bf q}_1'^2 }
\right)^{\frac{\bar \alpha}{2}\left(1-\frac{\zeta_2}{2} {\bar \alpha}\right) {\rm Y}} e^{\frac{3}{4} \zeta_3 {\bar \alpha}^2 {\rm Y}}
\Bigg\{\delta^{(2)} \left({\bf q}_1-{\bf q}_2\right) \nonumber\\
&&\hspace{-3.5cm}+\sum_{n=1}^\infty \prod_{i=1}^n \,  \Bigg[ \int d^2 {\bf k}_i 
{{\bar \alpha} \over 4} \left(1-\frac{\zeta_2}{2} {\bar \alpha}\right) 
 {\theta \left({\bf k}_i^2 - \lambda^2\right) \over \pi {\bf k}_i^2}
\Bigg(1+{\left({\bf q}_1'+\sum_{l=1}^{i-1}{\bf k}_l\right)^2 ({\bf q}_1+\sum_{l=1}^{i}{\bf k}_l)^2 - {\bf q}^2 {\bf k}_i^2 \over ({\bf q}_1'+\sum_{l=1}^{i}{\bf k}_l)^2 \left({\bf q}_1+\sum_{l=1}^{i-1}{\bf k}_l\right)^2}\Bigg) \nonumber\\
&&+ \Phi \left({\bf q}_1+\sum_{l=1}^{i-1}{\bf k}_l,{\bf q}_1+\sum_{l=1}^{i}{\bf k}_l\right)\Bigg]
\delta^{(2)} \left({\bf q}_1+\sum_{l=1}^n {\bf k}_l-{\bf q}_2\right)  \nonumber\\
&&\hspace{-3cm} \times\int_0^{y_{i-1}} \hspace{-.3cm} d y_i 
\left(\left({\bf q}_1+\sum_{l=1}^{i-1}{\bf k}_l\right)^2 \over 
\left({\bf q}_1+\sum_{l=1}^{i}{\bf k}_l\right)^2 \right)^{1+ {{\bar \alpha} y_i \over 2}\left(1-\frac{\zeta_2}{2} {\bar \alpha}\right) }
\left(\left({\bf q}_1'+\sum_{l=1}^{i-1}{\bf k}_l\right)^2 \over 
\left({\bf q}_1'+\sum_{l=1}^{i}{\bf k}_l\right)^2 \right)^{{ {\bar \alpha} y_i \over 2}\left(1-\frac{\zeta_2}{2} {\bar \alpha}\right) }\Bigg\}, 
\end{eqnarray}
where $y_0 \equiv Y$.

It is very difficult to provide an analytic representation of this Green function. However, it is possible to solve it 
in the large $q$ limit since there exists an asymptotic conformal invariance in that 
region~\cite{Bartels:2008ce,Bartels:2008sc,arXiv:1111.0782}. In the present paper we use advanced Monte Carlo 
integration techniques to study the gluon Green function. We have explicitly checked that our numerical solution 
to the exact equation agrees at large $q$ with the results found in~\cite{Bartels:2008ce,Bartels:2008sc,arXiv:1111.0782}. For this we have taken the large momentum transfer limit of the NLO BFKL kernel, 
which agrees with that in~\cite{arXiv:1111.0782}, and write the equation in the form
\begin{eqnarray}
\Bigg\{\omega + \frac{\bar \alpha}{2} \left(1-\frac{\zeta_2}{2} {\bar \alpha}\right)
\log{\left( {  {\bf q}_1^2   \over  \lambda^2}\right)} 
-\frac{3}{4} \zeta_3 {\bar \alpha}^2
\Bigg\} 
{\cal F}_\omega \left({\bf q}_1,{\bf q}_2;{\bf q}\right) &=&
\delta^{(2)} \left({\bf q}_1-{\bf q}_2\right)\nonumber\\
&&\hspace{-9.5cm}+\int d^2 {\bf k} \Bigg\{
 {{\bar \alpha}_s \over 4} \left(1-\frac{\zeta_2}{2} {\bar \alpha}\right) 
 {\theta \left({\bf k}^2 - \lambda^2\right) \over \pi {\bf k}^2}
\Bigg(1+{({\bf q}_1+{\bf k})^2 -  {\bf k}^2 \over  {\bf q}_1^2}\Bigg) 
+ \Phi \left({\bf q}_1,{\bf q}_1 + {\bf k} \right)\Bigg\} \nonumber\\
&&\hspace{-5cm}\times {{\bf q}_1^2 \over \left({\bf q}_1 + {\bf k}\right)^2}
{\cal F}_\omega \left({\bf q}_1+{\bf k},{\bf q}_2;{\bf q}\right), 
\end{eqnarray}
with a much simpler function
\begin{eqnarray}
\Phi \left({\bf q}_1,{\bf q}_1 + {\bf k} \right) &=& {{\bar \alpha}^2 \over 32 \pi} {1 \over {\bf q}_1^2 }
\Bigg\{ \left(\frac{1}{2}   - { \left({\bf q}_1^2  +  \left({\bf q}_1 + {\bf k}\right)^2\right) \over {\bf k}^2} \right)
\ln^2{\left({{\bf q}_1^2 \over \left({\bf q}_1 + {\bf k}\right)^2} \right)} \nonumber\\
&&\hspace{-2cm}
- \frac{1}{2}  { \left({\bf q}_1^2  -   \left({\bf q}_1 + {\bf k}\right)^2\right) \over {\bf k}^2} 
\ln{\left({{\bf q}_1^2 \over \left({\bf q}_1 + {\bf k}\right)^2} \right)}
\ln{\left({{\bf q}_1^2  \left({\bf q}_1 + {\bf k}\right)^2  \over   {\bf k}^4}\right)}  \nonumber\\
&&\hspace{-2.5cm} + \Bigg[{\bf k}^2 - 2 {\bf q}_1^2 - 2 \left({\bf q}_1 + {\bf k}\right)^2  
+ { \left({\bf q}_1^2   -   \left({\bf q}_1 + {\bf k}\right)^2\right)^2 \over {\bf k}^2}  \Bigg] 
{\cal I} \left({\bf q}_1^2, \left({\bf q}_1 + {\bf k}\right)^2, {\bf k}^2\right)\Bigg\}.
\end{eqnarray}
We present the results obtained from the numerical implementation of these representations in the following 
section. 

\section{Numerical results}

\begin{figure}[htbp]
\begin{center}
\includegraphics[width=12cm,angle=0]{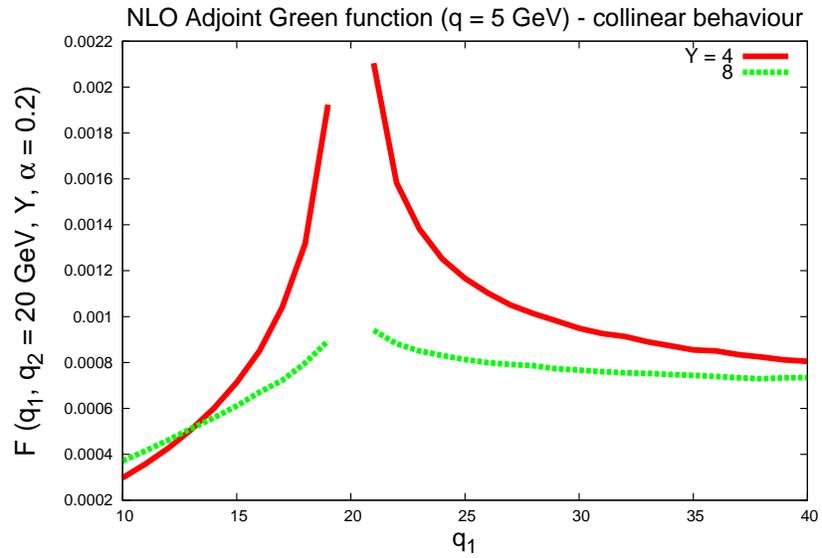}\\
\includegraphics[width=12cm,angle=0]{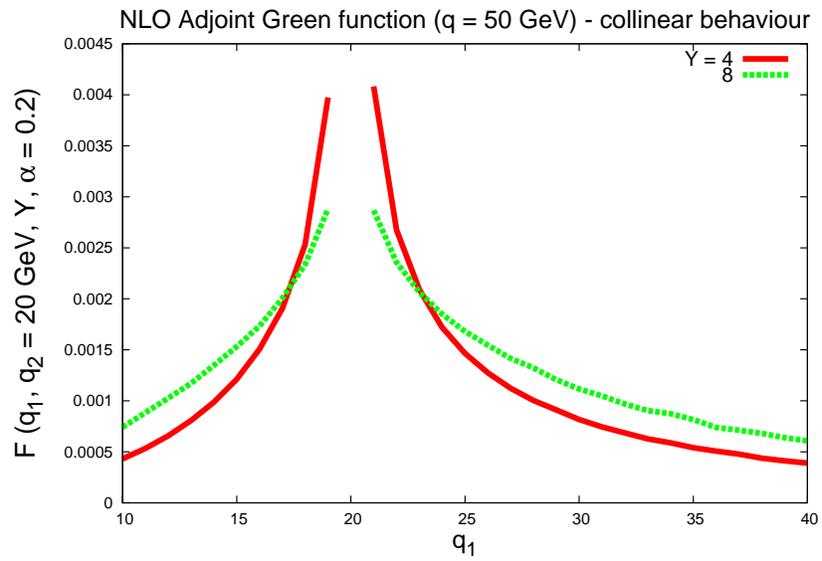}
\caption{Collinear behaviour of the adjoint gluon Green function at small (Top) and large (Bottom) momentum 
transfer.}
\label{Collinear}  
\end{center}
\end{figure}

We first present our solution for the gluon Green function in the collinear/anti-collinear regions where we fix $q_2=20$~GeV and 
vary $q_1$ from 10 to 40 GeV. This is shown, for $Y=4,8$, in Fig.~\ref{Collinear}. The discontinuity seen at 
$q_1 = q_2$ corresponds to the delta function initial condition chosen for the Green function. Its width is smaller at 
larger rapidities since a larger number of iterations of the kernel are needed to reach a convergent result, rapidly screening the information stemming from the initial condition. Hence, 
the weight in the full solution of the initial term is reduced. 

We have 
checked that all of our results are $\lambda$ independent for small $\lambda$, in agreement with the 
infrared finiteness of ${\cal F}$.  The qualitative collinear behaviour of the solution at large $q$ ($q=50$ GeV in 
Fig.~\ref{Collinear} (bottom)) is not very different to the one we previously found at LO in Ref.~\cite{Chachamis:2011nz}. 
In the present work we have also investigated the low $q$ region and found a flatter Green function for 
large $q_1$ than in the case of large $q$. It would be interesting to investigate how this is related to the lack of 
$SL(2,C)$ invariance at low $q$ and how it might affect the analytic calculation of the anomalous dimensions. 

\begin{figure}[htbp]
\begin{center}
\includegraphics[width=12cm,angle=0]{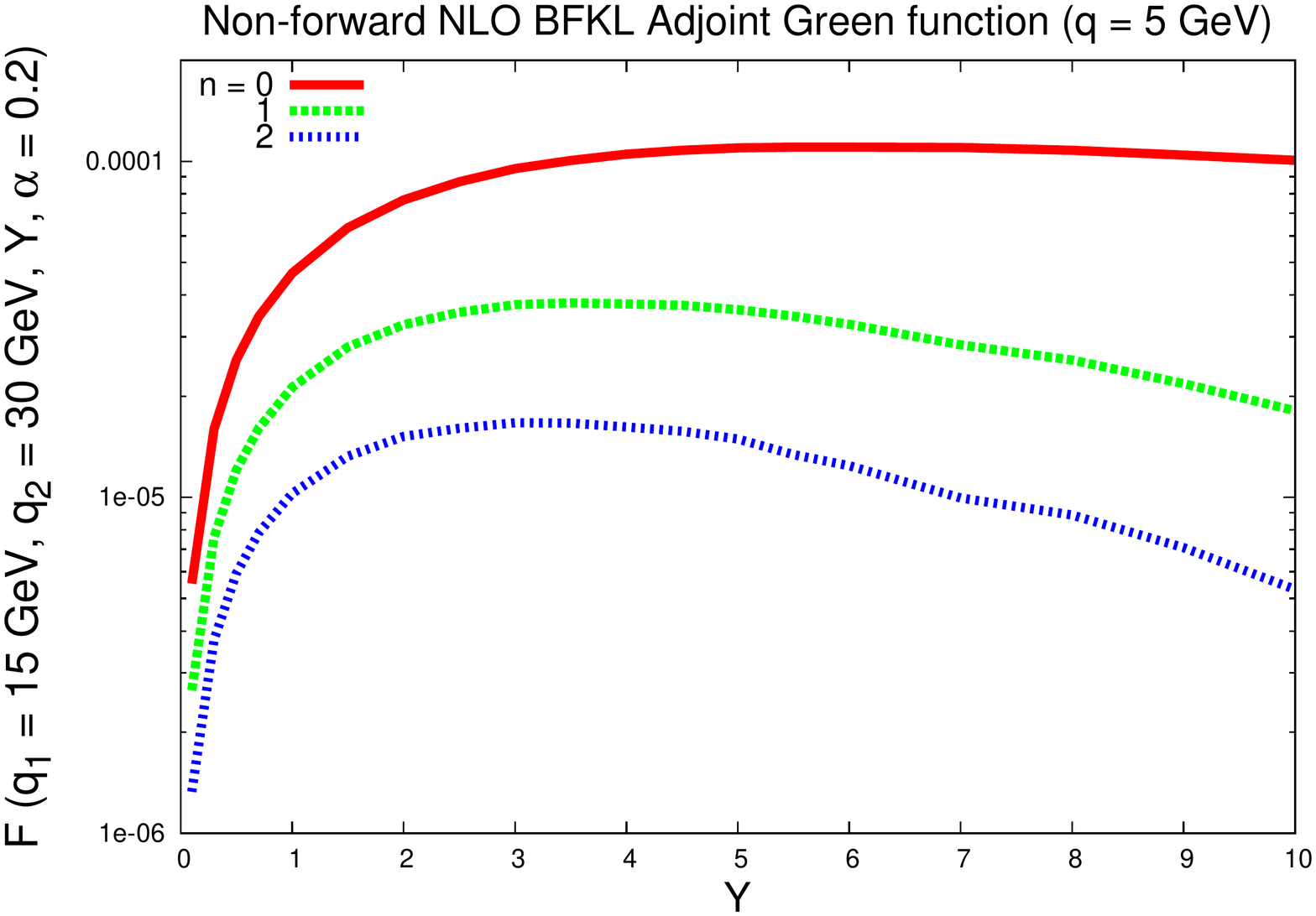}\\
\includegraphics[width=12cm,angle=0]{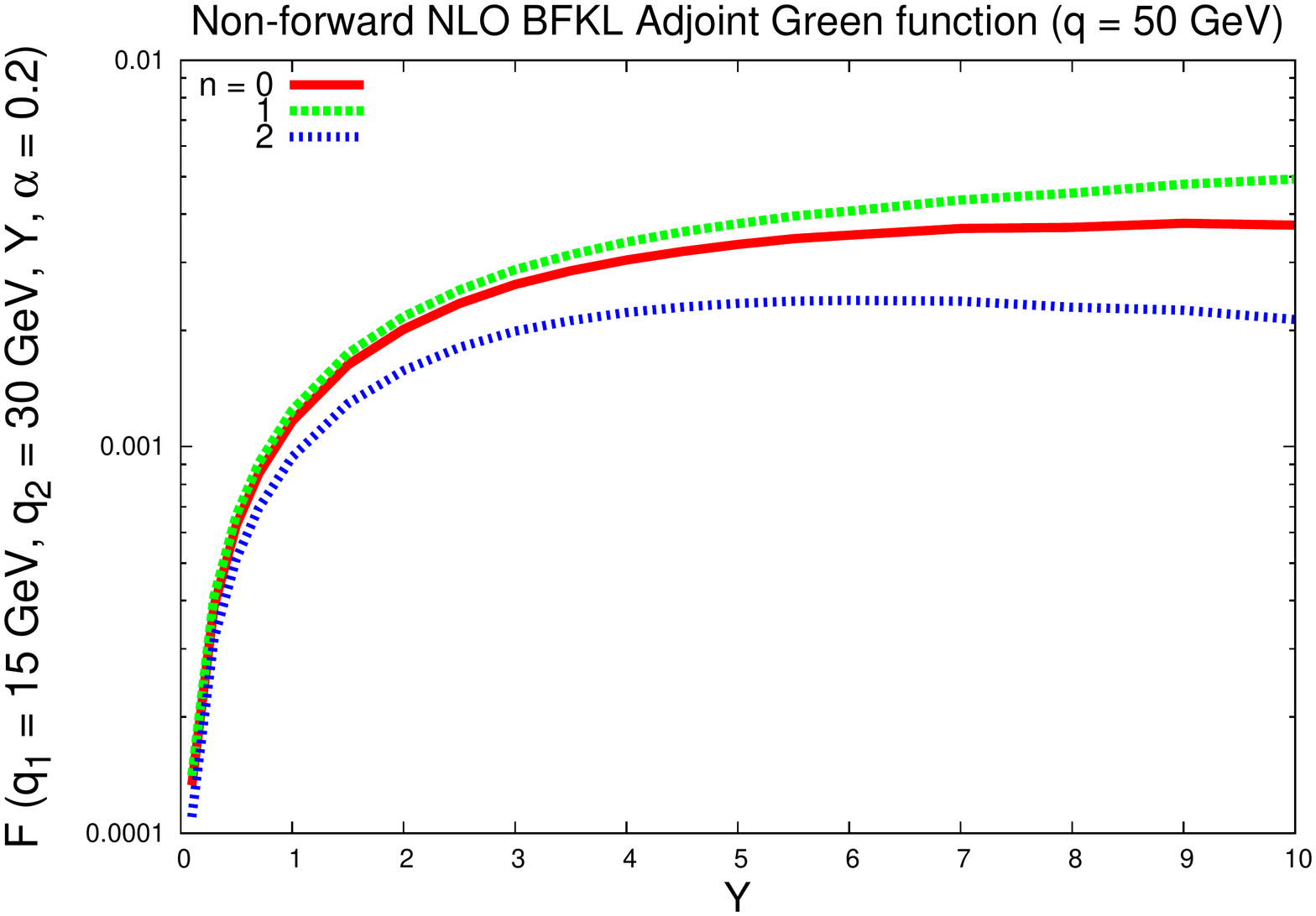}
\caption{Small momentum transfer (Top) and large momentum transfer (Bottom) evolution with energy of the different Fourier components of the NLO gluon Green function}
\label{Angles}  
\end{center}
\end{figure}

In studies of the BFKL gluon Green function it is always interesting to investigate its expansion in Fourier 
components in the azimuthal angle between the two transverse momenta ${\bf q}_1$ and ${\bf q}_2$. We have 
performed this analysis and briefly present some of the results in Fig.~\ref{Angles}. For large $q$ (bottom of 
the figure) we find qualitatively the same behaviour as at LO, with the only rising with energy component being 
the $n=1$ one. This is a common feature of the adjoint solution at large momentum transfer. We find an interesting 
change in this trend when $q$ is small. In this case, as it can be seen in Fig.~\ref{Angles} (Top) for $q=5$ GeV, 
the dominant Fourier component is that with $n=0$, with all the other components 
decreasing at high energies. Again, this should indicate the departure from conformal invariance. 

This is a brief presentation of our numerical studies. It will be very interesting to numerically integrate this solution
with the corresponding impact factors and extract information for the MHV amplitudes, to complete the already 
available studies in the literature.

\section{Conclusions and scope}

We have presented the exact solution of the BFKL equation in the adjoint 
representation for the ${\cal N}=4$ SUSY theory at NLO accuracy.  We have found agreement with the approximations to this solution in the case of large momentum transfer discussed in~\cite{Bartels:2008ce,Bartels:2008sc,arXiv:1111.0782}. The NLO non-forward BFKL gluon 
Green function plays a fundamental role in the construction of the ``finite remainder 
function" of MHV and planar amplitudes~\cite{Bartels:2008ce,Bartels:2008sc,arXiv:1111.0782}. It  
has been investigated in terms of energy growth, collinear limits and azimuthal angle 
behaviour. We have shown that the factorization of infrared divergencies is complete, 
generating an infrared finite gluon Green function. Our next task will be to convolute it with the corresponding impact factors to generate predictions for the scattering amplitudes.
\\
\\
\\
{\bf \large Acknowledgements}\\
We thank Victor Fadin and Lev Lipatov for discussions. G. C. thanks the Department of Theoretical Physics at the Aut{\'o}noma University of Madrid and the ``Instituto de F{\'\i}sica Te{\' o}rica 
UAM / CSIC" for their hospitality. We thank the CERN PH-TH Unit where the first stages of this work took place. 
Research partially supported by the European Comission under contract LHCPhenoNet (PITN-GA-2010-264564), 
the Comunidad de Madrid through Proyecto HEPHACOS ESP-1473, and MICINN (FPA2010-17747).

\end{document}